\documentclass[]{pasj01} %図を表示する
\Received{$\langle$03--Aug--2022$\rangle$}
\Accepted{$\langle$05--Dec--2022$\rangle$}
\Published{$\langle$publication date$\rangle$}

\usepackage{longtable}
\usepackage{graphicx}
\usepackage{txfonts}
\usepackage{natbib}
\usepackage{subfloat}
\usepackage{lineno}
\def\la{\ifmmode{\,\lesssim\,}\else$\,\lesssim$\,\fi}
\def\ga{\ifmmode{\,\gtrsim\,}\else$\,\gtrsim$\,\fi}

\begin{document}

\title{An expanding ring of the hypercompact HII region W49N:A2}

\author{Ryosuke Miyawaki$^1$, Masahiko Hayashi$^{2,3}$, and Tetsuo Hasegawa$^2$}%
% ORCID
%Ryosuke Miyawaki: https://orcid.org/0000-0001-5259-4080
%Masahiko Hayashi: https://orcid.org/0000-0002-4790-7940
%Tetsuo Hasegawa: https://orcid.org/0000-0003-1853-0184 

%
\altaffiltext{1}{College of Arts and Sciences, J.F. Oberlin University, Machida, Tokyo 194-0294, Japan}
\email{miyawaki@obirin.ac.jp}
\altaffiltext{2}{National Astronomical Observatory of Japan, 
2-21-1 Osawa, Mitaka, Tokyo 181-8588, Japan}
\altaffiltext{3}{JSPS Bonn Office, Ahrstr. 58, 53175 Bonn, Germany}

%stars: winds, outflows stars: winds, outflows ISM: clouds ISM: jets and outflows %検討
\maketitle

%: ABSTRACT
\begin{abstract}\label{ABSTRACT}
We present 250~GHz continuum and H29$\alpha$ line data toward W49N:A2, a hypercompact HII region ionized by an O9 star.
The data obtained with ALMA at a resolution of $\sim$0\farcs05 (600~au) confirmed the presence of an ionized ring with a radius of $\sim$700~au inclined by $\sim$50\degree\ (0\degree\ for pole-on).
It has a width of $\sim$1000~au and is relatively flat with a scale height of less than several hundred au.
The tilted ring, or the apparent ellipse, has a prominent velocity difference between its NW and SE ridges along the minor axis, suggesting that it is expanding in the equatorial plane at a velocity of 13.2~km\,s$^{-1}$.
The ring also shows a hint of rotation at 2.7~km\,s$^{-1}$, which is significantly (2.5\,$\sigma$) smaller than the Kepler velocity of 5.2~km\,s$^{-1}$ at its radius around the 20~M$_\odot$ star.
This can be interpreted that the ring gas has been transported from the radius of $\sim$170~au by conserving its original specific angular momentum that it had there.
The ionized ring may thus be a remnant of the accretion disk that fed the O9 star, whose radiation or magnetic activities became so strong that the disk accretion was reversed due to the intense thermal or magneto-hydrodynamic pressure around the star.
The data has revealed a rare example of how a massive star terminates its accretion at the end of its formation, transforming a hypercompact HII region into an ultracompact HII region. 
\end{abstract}
%\linenumbers

%: INTRODUCTION
\section{Introduction}\label{INTRODUCTION}
Hypercompact HII (HCHII) regions are characterized by their small sizes ($<$0.05 pc or 10,000~au), high electron densities ($n_{\rm e} >$ 10$^6$ cm$^{-3}$), large emission measures ($>$10$^8$ pc cm$^{-6}$), and broad radio recombination line widths ($\Delta V >$ 40~km\,s$^{-1}$) \citep{Sewilo2004, Kurtz2005, Murphy2010}.
They are considered to precede the phase of ultracompact HII (UCHII) regions in massive star formation.

Photoevaporating accretion disk models may explain the characteristics of HCHII regions \citep{Hollenbach1994, Lizano1996, Keto2007}, in which inevitable outflows may be responsible for the broad radio recombination lines \citep[e.g.,][]{Tanaka2016}.
Alternatively, the broad line width may also be explained by the pressure broadening caused by the high-density ionized gas in HCHII regions \citep[e.g.,][]{Brocklehurst1972, Kurtz2005}.
Numerical calculations show that a small quasi-spherical HII region first forms within the accretion flow \citep{Keto2007}.
The initial HII region is gravitationally trapped within the Bondi-Parker radius, while the accretion may proceed through the HII region \citep{Keto2002}.
As the ionizing flux increases, the HII region expands driven by the ionized gas pressure, transitioning to an ionized bipolar outflow, and eventually, as the outflow opening angle increases, the accretion flow is confined to a thin disk.
The photoionization of the bipolar low-density cavities pushes mass from high latitudes toward the disk.
This feedback yields an increase in disk mass and hence stellar accretion rate in the last stage of massive star formation \citep{Kuiper2018}.

W49N:A is one of the UCHII regions cataloged by \citet{DePree1997} in their 8.3~GHz map. 
Higher resolution observations at 45~GHz have resolved it into two sources A1 and A2 \citep{DePree2000}, with A2 showing a distinct ring located at the center of the oppositely directed, edge-brightened bipolar lobes of ionized gas \citep{DePree2020}.
W49N:A2 is ionized by a Lyman continuum photon luminosity equivalent to an O9 ZAMS star and has the total ionized gas mass of 0.12~M$_\odot$ \citep{DePree2020}.
In this paper, we present the 250~GHz continuum and H29$\alpha$ line data for W49N:A2 and examine its kinematics.
We assume the distance to W49N to be 11.11$^{+0.79}_{-0.69}$~kpc \citep{Zhang2013} throughout this paper.

%: OBSERVATIONS
\section{ALMA archival data}\label{OBSERVATIONS}
We used Atacama Large Millimeter/submillimeter Array (ALMA) archival data (\#2018.1.00520.S: PI D. Wilner) for the study of protostars in W49N. 
The observations were performed from July 2019 to September 2019 using 43--48 12-m antennas. 
The shortest and longest baselines for the 12-m antennas were 38.4\,m and 3637.7\,m, respectively.
Flux, bandpass, pointing, and phase calibrations were carried out with J1905+0952, J1907+0907, and J1908+1201. 

Four spectral windows (240.76--242.63 GHz, 242.61--244.49~GHz, 255.86--257.74~GHz, and 257.70--259.57~GHz), each covering a $~2$\,GHz bandwidth, were set up to observe the target source W49N.
Image analysis was done using the CASA software \citep{CASA2022}.
We separated the continuum and line emissions by fitting a linear baseline to line free channels of each spectral window using the `uvcontsub' task of CASA.
The continuum data at the effective frequency of 250\,GHz was then made by averaging the line-free intensities of all four windows.

For the H29$\alpha$ line data, we set up a data cube with a spectral resolution of 2~kms$^{-1}$.
Although the frequency resolution (channel width) was 1128.91\,kHz ($\sim$1.31~km\,s$^{-1}$), there was variation in frequency to channel relation between datasets obtained on different days.
% Rounding
We thus proportionally distributed the flux received in an original spectral channel into the nearest 2~kms$^{-1}$  bins by setting the width parameter of the `tclean' task to be 2~kms$^{-1}$ to compensate for the variation.
%Data with poor signal-to-noise ratio were not used.
We analyzed three sets of data 3.1 hours long in total, but the second set showed a large noise level probably due to errors in the calibration process, and was not used.

The phase center was $\alpha$(ICRS) = 19$^{\rm h}$10${\rm ^m}$13\,\fs\,300 and $\delta$(ICRS) =09\degree06$'$14$\,\farcs$00 with a primary beam field of view (HPBW) of 23\farcs3.
We set the parameters of the `tclean' task as weighting=`briggs' and robust=`0.5'.
This setting creates a PSF that smoothly varies between natural and uniform weighting based on the signal-to-noise ratio of the measurements and a tunable parameter that defines the noise threshold.
The synthesized beam size of the continuum image was 0\farcs046 $\times$ 0\farcs032 (PA=$70.3\degree$) at the effective frequency of 250~GHz, and that of the H29$\alpha$ recombination line image was 0\farcs052$\times$ 0\farcs033 (PA=$58.0\degree$) at its frequency of 256\,GHz.
The resultant noise levels were 0.7\,mJy\,beam$^{-1}$ and 2.0\,mJy\,beam$^{-1}$ for the continuum and line maps, respectively.

%: RESULTS & DISCUSSION
\section{Results and discussion}\label{RESULTSandDISCUSSION}

\subsection{Continuum and H29$\alpha$ Rings}\label{Cont+H29α}

\begin{figure}[tb]
\includegraphics[bb= 0 0 400 410, scale=0.58]{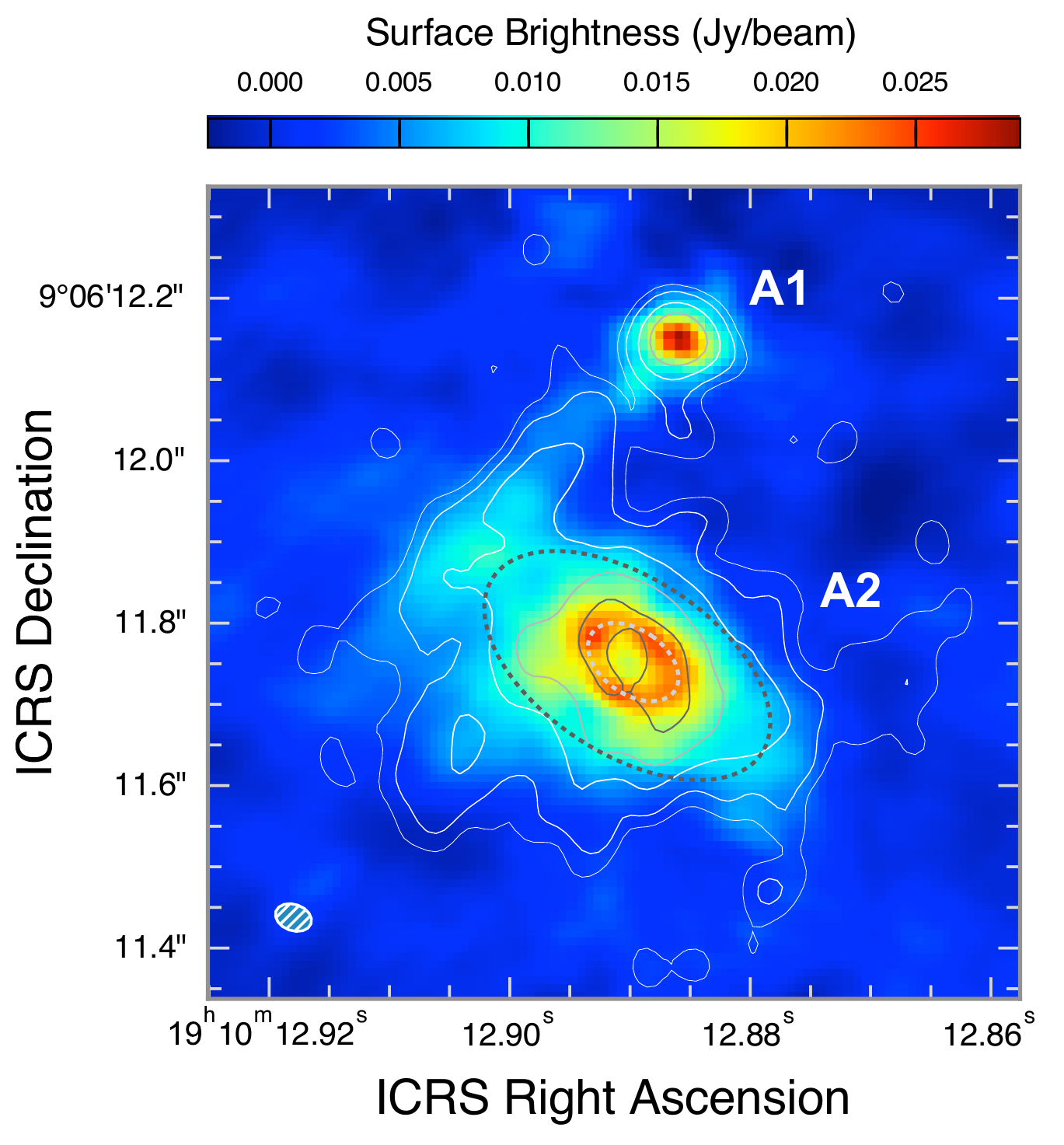}
\caption{The 250\,GHz continuum map superposed on the contours of the 45\,GHz continuum emission \citep{DePree2000} at 1, 2, 5, 10, and 20 mJy\,beam$^{-1}$. 
The two dotted ellipses show the ring (white) and the FWHM size of the entire emission nebula (black).
}
\label{Fig01}
\end{figure}

Figure~\ref{Fig01} shows the 250\,GHz continuum image of W49N:A2, with A1 also seen at its $\sim$0\farcs4 north.
A2 exhibits an extended emission elongated from the NE to SW with a central dip surrounded by a ridge of elliptical emission.
This feature was first noted by \citet{DePree2000} in their 45~GHz continuum map, which is shown by contours in Figure~\ref{Fig01} for comparison.\footnote{The 45~GHz continuum map is shifted by ($\Delta\alpha$, $\Delta\delta$)=($-$0\farcs019, $-$0\farcs065) from its original coordinates so that the peak position of A1 coincides with that of the 250~GHz map to compensate the unknown positional displacement described by \citet{Rodriguez2020}. 
We also applied the task `phaseshift' to CLEAN the data but did not obtain any different results.}
This kind of annular morphology is relatively common in UCHII regions and is interpreted as a ring or a shell, which may be distinguishable from the observed contrast ratio between the flux density of the shell peak and that of the cavity \citep{Turner1984}.
For the 45~GHz image of W49N:A2, the ratio is {\it inconsistent} with a shell morphology, and the observed elliptical structure is interpreted as a ring or a flattened disk with a central gap of diameter $\sim$0\farcs08 ($\sim$880~au) \citep{DePree2000, DePree2020}.
The major axis of the ellipse is perpendicular to the axis of the ionized bipolar outflow seen as an edge-brightened double-lobed structure in the 8.3~GHz maps \citep{DePree1997, DePree2020}, which also suggests that the observed elliptical structure is a tilted equatorial ring perpendicular to the outflow.
We, therefore, assume in this paper that the 250~GHz elliptical emission represents a tilted ring, which is intrinsically circular on the equatorial plane of the star. 
The ring is supposed to be geometrically and kinematically symmetric with respect to the polar axis of the star (rotational symmetry) and its equatorial plane (plane symmetry).

%:%%%%% Fig01 250GHz continuum %%%%%
%%%%%  250GHz continuum %%%%%

%:%%%%% Table 1 Properties of Disk  %%%%%
\begin{table*}[tbh]
\caption{Observed parameters}

\label{Table01}
\begin{center}
\scalebox{0.5}[0.5]

\begin{tabular}{llllllll}

\hline\hline
%Properties&  &  \\
%\hline
%Peak Position&  &  \\
& 250~GHz continuum & H29$\alpha$ line & 45~GHz continum$^*$\\
\hline

\multicolumn{3}{l}{Center Position}\\
\ \ $\alpha$(ICRS) 19$^{\rm h}$10${\rm ^m}$+ &
12\fs8921  $\pm$0\fs0003 &
12\fs8927  $\pm$0\fs0005  &
12\fs8929  $\pm$0\fs0001 \\

\ \ $\delta$(ICRS) 09\degree06$'$+& 
11$\farcs$756 $\pm$0$\farcs$004  & 
11$\farcs$781  $\pm$0$\farcs$008 & 
11$\farcs$837 $\pm$0$\farcs$002\\

Beam Size (position angle)
&0$\farcs$046 $\times$ 0$\farcs$032 (70.3\degree)
&0$\farcs$052 $\times$ 0$\farcs$033 (61.5\degree)
&0$\farcs$055 $\times$ 0$\farcs$044 (8.6\degree )\\

\multicolumn{3}{l}{Image Component Size$^\dagger$}\\
\ \ Major axis FWHM 
& 400 $\pm$20 (mas)
& 510 $\pm$30 (mas)
& 270 $\pm$6 (mas) \\

\ \ Minor axis FWHM 
& 220 $\pm$ 10 (mas)
& 310 $\pm$ 20 (mas)
& 182 $\pm$  4 (mas) \\

\ \ Position Angle
& 58\degree $\pm$2\degree 
& 40\degree $\pm$4\degree
& 53\degree $\pm$3\degree\\

\ \ Inclination Angle
& 57\degree $\pm$2\degree 
& 53\degree $\pm$3\degree
& 48\degree $\pm$2\degree\\

Peak Brightness 
& 18.3 $\pm$0.1  (mJy\,beam$^{-1}$)
& 1.48 $\pm$0.07 (Jy\,beam$^{-1}$ km\,s$^{-1}$) 
& 22.2 $\pm$0.5  (mJy\,beam$^{-1}$) \\

Total Flux Density
& 1.11 $\pm$0.03 (Jy)
& 152 $\pm$7 (Jy km\,s$^{-1}$)
& 0.47 $\pm$0.01 (Jy) \\

\multicolumn{3}{l}{Ring Size$^\ddagger$}\\
\ \ Major  Axis (PA=55\degree)& 
124$\pm$4 (mas) &
116$\pm$10 (mas) &
117$\pm$5 (mas) \\

\ \  Minor Axis  (PA=145\degree)&
79$\pm$1 (mas) &
82 $\pm$3 (mas) &
96$\pm$4 (mas) \\

\ \ Inclination Angle
& 50\degree $\pm$2\degree 
& 45\degree $\pm$6\degree
& 35\degree $\pm$5\degree\\

\multicolumn{3}{l}{Ring Width$^\dagger$$^\ddagger$}\\
\ \ Major Axis FWHM ($2a$)& 
90$\pm$12 (mas) &
103$\pm$41 (mas) &
84$\pm$38 (mas) \\

\ \ Minor Axis FWHM ($2\beta$)& 
54$\pm$9 (mas) &
62$\pm$11 (mas) &
78$\pm$39 (mas) \\

$\cos^{-1}$($\beta/a$)
& 53\degree $\pm$9\degree 
& 53\degree $\pm$19\degree
& 22\degree $\pm$97\degree\\

\hline
\end{tabular}
\end{center}
%\\
\footnotetext{1}{$^*$\citet{DePree2000, DePree2020}. The center position shows their original coordinates (see footnote 1).}\\
\footnotetext{1}{$^\dagger$Beam deconvolved FWHM}\\
\footnotetext{2}{$^\ddagger$Measured by fitting a double Gaussian to each spatial profile.}
\end{table*}

%%%%% Table 1%%%%%

Table~\ref{Table01} lists the observed parameters of W49N:A2 derived from the 2D Gaussian fitting tool of CASA.
The total flux density of A2 at 250~GHz is 1.11~Jy, which is compared with its 45~GHz flux density of 0.47~Jy to give a spectral index $\alpha$ =~0.5 ($F_\nu\propto\nu^\alpha$).
Preliminary measurements using additional archival data resulted in the flux densities of 1.3$\pm$0.3~Jy, 2.7$\pm$0.2~Jy, 1.3$\pm$0.2~Jy, and 0.59$\pm$0.09~Jy at 100~GHz, 230~GHz, 360~GHz, and 450~GHz, respectively, including both A2 and A1.\footnote{ALMA archival data: 100GHz (\#2018.1.00589.S PI: R. Galv\'{a}n-Madrid), 230GHz (\#2016.1.00620.S PI: A. Ginsburg), 360GHz (\#2015.1.01535.S PI: B.~R. Alejandro and \#2017.1.00318.S PI: R. Galv\'{a}n-Madrid), 450GHz (\#2017.1.01499.S PI: F. Xiaoting)}
If we use all these data points, we obtain $\alpha$ = 0.4 $\pm$0.7, indicating that the source has a flat or slowly rising spectrum.
The flux density distribution does not show any steeply ($\alpha>2$) rising tendency toward submillimeter wavelengths, suggesting a negligible contribution from thermal dust emission at these frequencies.
We examined any molecular line emission directly toward the A2 nebulocity in the current archive data, finding no molecular lines significantly detected within 200 mas in radius, inside which the ionized ring is located. 
In particular, the lines of CH$_{3}$CN ($J_{K}$ = 14$_{K} - 13_{K}$), which would be a dominant molecular emission in the band if molecular gas existed, show very little ($<3\sigma$) emission there.
In fact, the CH$_{3}$CN  emission was detected around A2 at radii larger than 200 mas, corresponding to the outside of the main ionized nebula A2.
This tendency is the same for emissions from other molecular lines such as $^{13}$CO or SO.
In other words, there is a hole of molecular emission toward A2, suggesting that no neutral disk is associated with the ionized ring inside the emission nebula A2.

The 250~GHz emission has the major and minor axis lengths of 400~mas $\times$ 220~mas (deconvolved FWHM), respectively, corresponding to 4400~au $\times$ 2400~au, with a major axis position angle of 58\degree.
This gives an inclination angle of 57\degree$\pm$2\degree\ (0\degree\ for pole-on) if a geometrically thin circular structure is assumed.
The distance between the two ridges of the ring is 124$\pm$4~mas along the major axis, corresponding to 690~au, and 79$\pm$1~mas along the minor axis, when a double gaussian fit is applied to the spatial profiles. 
The minor to major axis ratio of the ring suggests an inclination of 50\degree$\pm$2\degree\ if we assume that the ring is intrinsically circular when viewed pole-on.

Figure~2 shows the moment~0 map of the H29$\alpha$ emission integrated over the velocity range from $-$20 to 60 km\,s$^{-1}$ superposed on the 250~GHz continuum contours.
The line emission follows the continuum map well, showing the elliptical ring of 116~mas $\times$ 82~mas, which are the distances between the ridges along the major (PA=55\degree) and minor  (PA=145\degree) axes, respectively.
The ring's axis lengths are consistent between the 250 GHz and H29$\alpha$ data within the errors.
We can thus naturally assume that the H29$\alpha$ emission represents the same ring as the 250~GHz continuum.
The minor to major axis ratio of the H29$\alpha$ ring gives the inclination angle of 45\degree$\pm$6\degree.

%:%%%%% Fig02 H29alpha moment0 with 250GHz continuum %%%%%
\begin{figure}[htb]
%\begin{figure}
\includegraphics[bb= 0 0 400 410, scale=0.58]{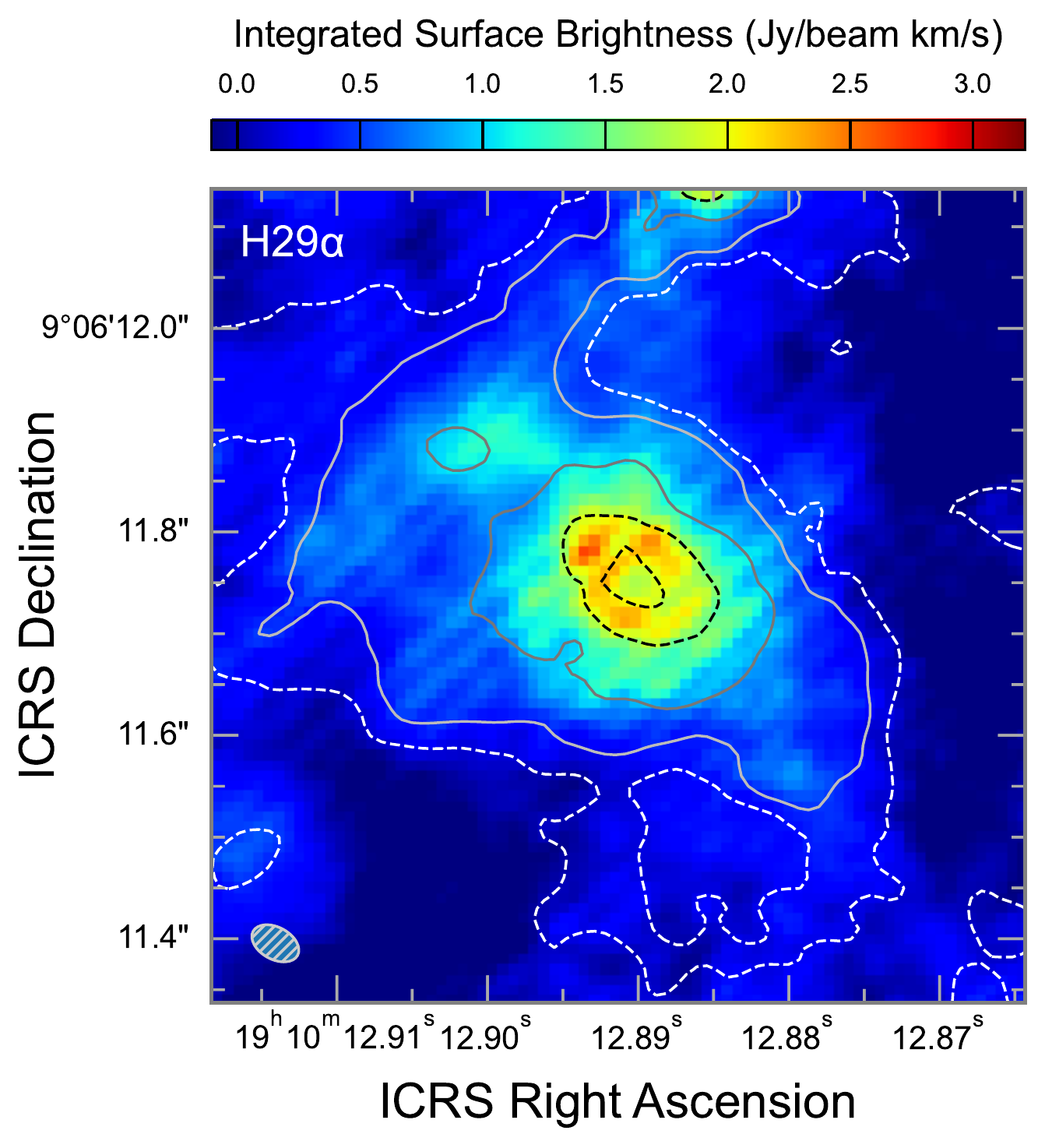}
\caption{Integrated intensity (moment~0) map of the H29$\alpha$ emission superposed on the 250\,GHz continuum contours at 1, 2, 5 and 10~\% of 194 mJy\,beam$^{-1}$. 
The integration range is from $V_\mathrm{LSR} = -$20 to 60~km\,s$^{-1}$.
}
\label{Fig02}
%\end{figure}
\end{figure}
%%%%% Fig02 H29alpha moment0 %%%%%

Table~\ref{Table01} also lists the beam deconvolved FWHM widths of the ring along the major and minor axes.
The widths along the ring's major axis, corresponding to its radial width on the equatorial plane, are 90 mas (1000 au) and 100 mas (1100 au) for the 250 GHz and H29$\alpha$, respectively, 
They are consistently larger than the widths of 54 mas and 60 mas along the minor axis for the 250 GHz and H29$\alpha$ rings, respectively.
This means that the ring has a smaller thickness than its radial width as we will discuss below.
This tendency was not confirmed for the 45 GHz ring.
This is because the 45 GHz ring does not show a clear elliptical shape \citep[see Fig. 2 (i) of][]{DePree2000}, and the double Gaussian fit to its spatial profiles along the major and minor axes gives large errors for the ring widths.
We will thus use only the 250 GHz and H29$\alpha$ data for the following discussion about the scale height.

To estimate the scale height of the ring, we assume a simple model that the ring has an elliptical cross-section with the semi-axis lengths of $a$ and $b$ along the ring's equatorial plane and polar axis, respectively.
The configuration is shown in Figure~3.
The ellipse on the left shows the tilted ring projected on the plane of the sky, while the ellipse on the right is a cross-section of the ring cut along the plane defined by the ring's polar and radial axes.
Under this model, the value $2a$ represents the non-projected radial width of the ring and $2b$ stands for its thickness, i.e., two times the scale height.
The value $2\beta$ is the apparent width of the projected ring along its minor axis.

The ring width $2a$ can be directly measured by the width of the projected ring along its major axis, while the value $2\beta$ can be obtained from the width of the projected ring along its minor axis.
The scale height $b$ is then estimated from the values of $a$ and $\beta$ using the geometrical relation $\beta^2=a^2\,\cos^2 i + b^2\,\sin^2 i$, where $i$ is the inclination angle of the ring ($i=0\degree$ for pole-on).
If, for example, the ring cross section was circular, i.e., $a=b$, we would obtain the trivial result $\beta=a$.
If the ring was flat, i.e., $b=0$, we would obtain another trivial result $\beta=a\,\cos i$.

%:%%%%% Fig03 ring %%%%%
\begin{figure*}[tb]
\includegraphics[bb= 70 180 450 470, scale=0.5]{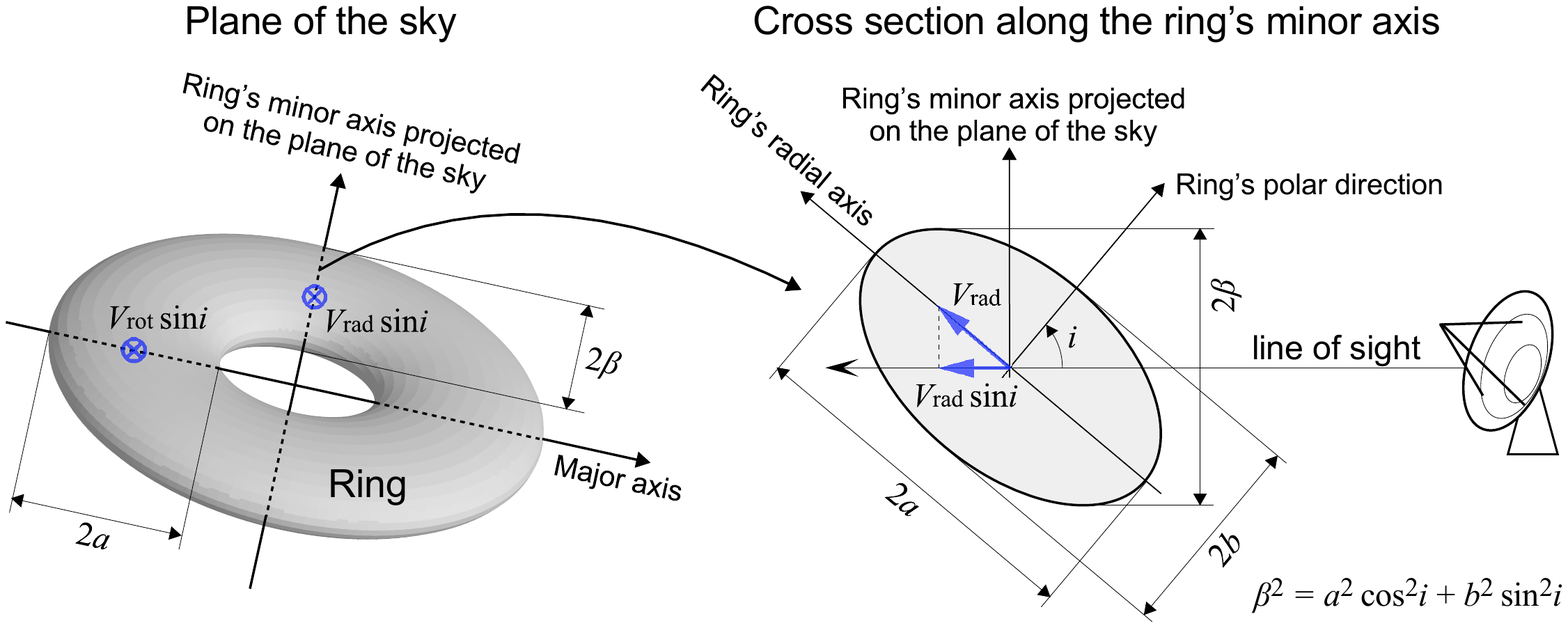}
\caption{A ring model with an elliptical cross-section. 
The observed width $2\beta$ of the ring along the minor axis on the plane of the sky is related to the elliptical semi-axis lengths $a$ and $b$ as $\beta^2=a^2\cos^2 i + b^2 \sin^2 i$, where $i$ is the inclination of the ring ($i=0$ for pole-on).
If the ring has a radial (expansion or contraction) velocity $V_{\rm rad}$ and a rotational velocity $V_{\rm rot}$ along its equatorial plane, the line of sight velocities toward the major and minor axes are $\pm V_{\rm ror}\sin i$ and $\pm V_{\rm rad}\sin i$, respectively.
}
\label{Fig03}
\end{figure*}
%%%%%  250GHz continuum %%%%

For the 250 GHz and H29$\alpha$ data in Table~\ref{Table01}, the ring widths $a$ and $\beta$ are relatively well defined.
The values of $\cos^{-1}(\beta/a)$ are consistent with the ring inclination angle of 50\degree\ and 45\degree, respectively, within the errors.
This means that the value $b=0$ is consistent with the observations, suggesting that the ring is relatively flat having a small scale height.
We estimate a 2$\sigma$ upper limit to the scale height $b$ by taking 2$\sigma$ larger and smaller values for $\beta$ and $a$, respectively.
The results are $b<$410~au and $b<$ 690~au for the 250 GHz and H29$\alpha$ rings, respectively.
As expected, the ring's scale height is smaller than its radial widths of $\sim$1000~au. 
It is also significantly smaller than the pressure scale height of $\sim$1700~au of an ionized gas disk of 10000~K around a 20~M$_\odot$ star at a radius of 690~au, implying that the ring is not a static structure and is being evaporated by the ionizing radiation from the central star.

% Scale height b = 410 au (37 mas) for β+2σ, a-2σ, and i=50.4° for 250 GHz
% Scale height b = 650 au (58 mas) for β+2σ, a-2σ, and i=45.０° for H29α

\subsection{Velocity structure of the ring}\label{Velocity Structure}

Figure~4 shows the velocity channel maps of the H29$\alpha$ line, together with a moment 1 (mean velocity) map on the bottom right panel.
It demonstrates a clear velocity shift along the minor axis of the elliptical ring.
The NW half of the ring is bright at the radial velocities less than 0~km\,s$^{-1}$, while the SE part is bright at velocities larger than 30~km\,s$^{-1}$.
Under our assumption that the ring is intrinsically circular and symmetric with respect to its polar axis and equatorial plane, the line of sight velocity toward the minor axis reflects only the radial motion of the tilted ring, but not its rotational motion (see Figure~3).
Thus the observed large velocity difference between the two ridges along the minor axis means that radial motion is dominant in the ring.
Whether the motion corresponds to expansion or contraction is not straightforward, and will be discussed in the next section (\S\ref{Expansion}).

%:%%%%% Fig04 H29alpha channel %%%%%
\begin{figure*}[tbhp]
\includegraphics[bb= 0 0 450 520, scale=0.42]{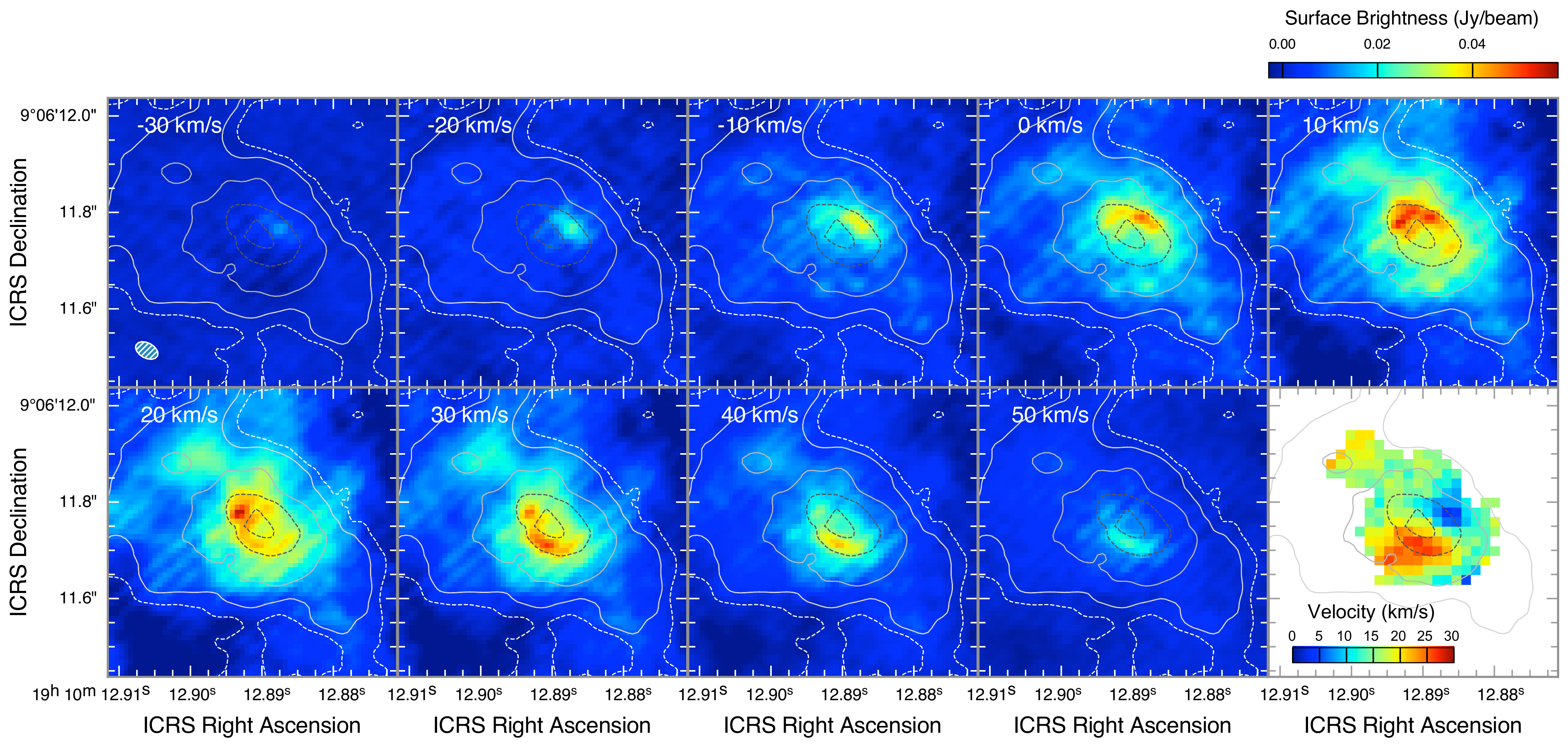}
\caption{Velocity channel maps of the H29$\alpha$ line superposed on the 250\,GHz continuum contours at 1, 2, 5 and 10~\% of 194 mJy\,beam$^{-1}$. 
The bottom right panel shows a mean velocity (moment 1) map.
}
\label{Fig04}
\end{figure*}
%%%%%  250GHz H29alpha channel  %%%%%

At $V_\mathrm{LSR}$ = $-$10 and 0~km\,s$^{-1}$, there is a hint that the NE side of the ring has stronger emission than the SW side.
At $V_\mathrm{LSR}$ = 40 and 50~km\,s$^{-1}$, the emission peak is slightly shifted to the west along the southern part of the ring with respect to the minor axis of symmetry.

\begin{figure}[htb]
\includegraphics[bb= 60 70 450 780, scale=0.85]{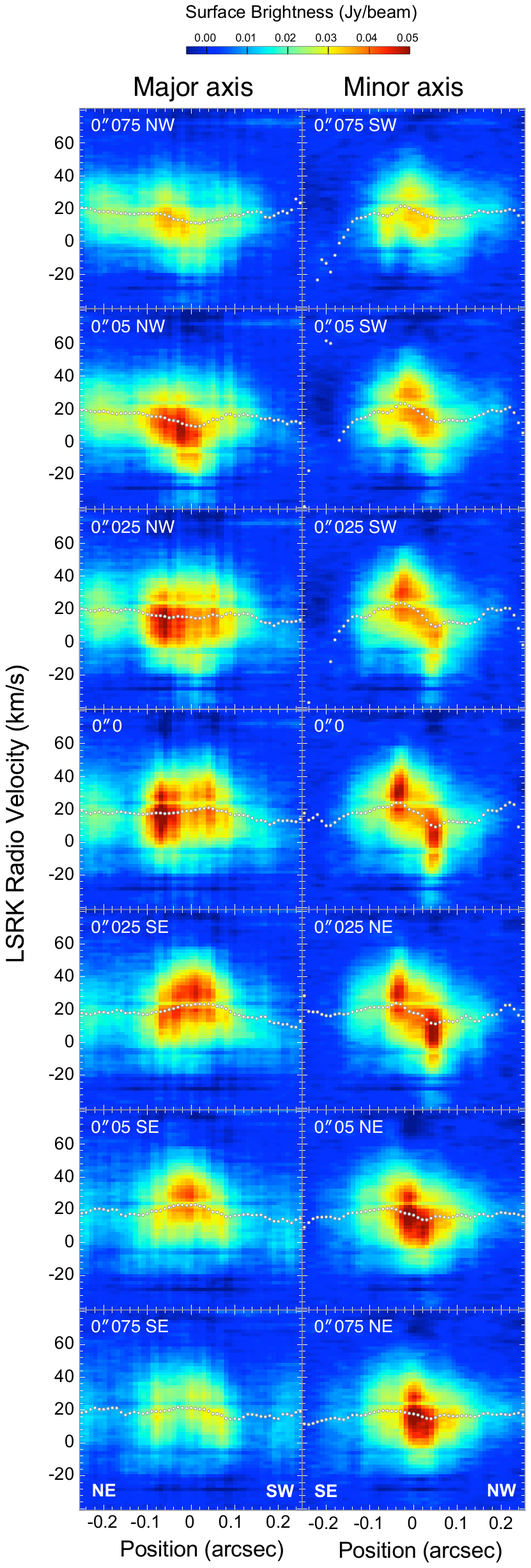}
\caption{Position-velocity diagrams of the H29$\alpha$ line along the strips parallel to the major (PA=55\degree, left) and minor (PA=145\degree, right) axes.
The brightness was averaged over three pixels (0\farcs03) perpendicular to each strip.
The offset of each strip from the major (left) or minor (right) axis is shown at the top left corner of each panel.
Small circles show mean velocities.
}
\label{Fig05}
\end{figure}

To examine this tendency in more detail, we show in Figure~\ref{Fig05} the position velocity diagrams along the major and minor axes.
The velocity difference across the minor axis is obvious (right panels).
The PV diagrams along the strips parallel to the major axis (left panels) show convex features on the SE side, and, alternatively, concave features on the opposite NW side.
Such convex/concave features in a PV diagram arise when the system has a systematic radial motion (expansion or contraction).
The entire ring thus has a consistent radial motion, either expansion or contraction\footnote{This conclusion is valid even if the ring, with its center at the exciting star, is tilted with respect to the equatorial plane. Thus, if we take an assumption that the ring is, for example, on the polar plane of the star, we derive the same conclusion that the ring has a radial motion. 
We, however, do not take such an assumption because it is simply unphysical: we do not find any reason that the inflowing or outflowing gas forms a regular ring on a plane significantly tilted with respect to the equatorial plane of the star, breaking the rotational symmetry about the polar axis.}

%:%%%%% Fig05 H29alpha PV  major axis %%%%%
%%%%%  Fig05 %%%%%

\begin{figure}[tb]
\includegraphics[bb= 50 260 400 720, scale=0.45]{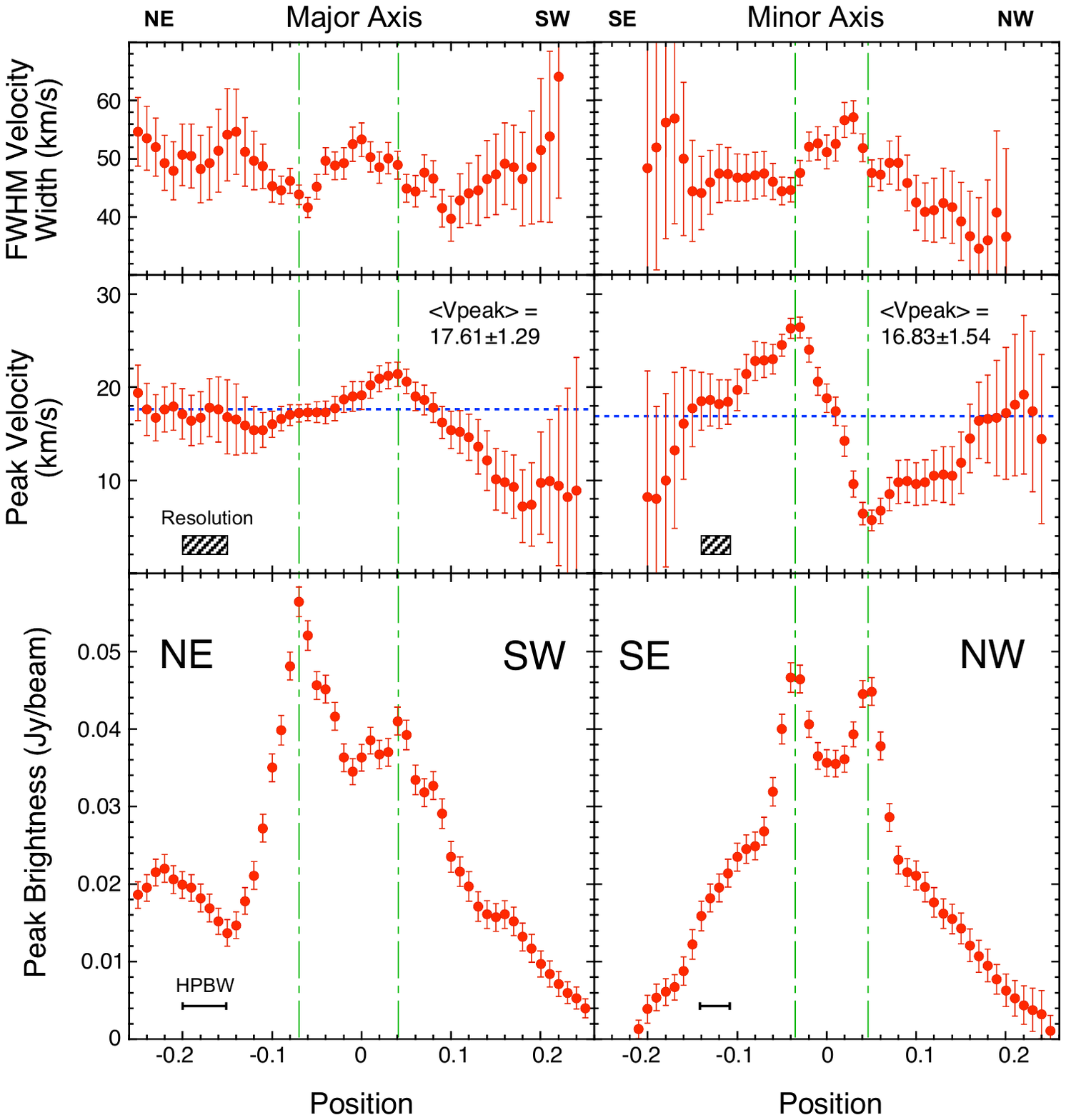}
\caption{Variations of peak brightness, peak velocity, and FWHM velocity width of the H29$\alpha$ line along the major (left) and minor (right) axes.
Data points with large errors are not plotted.
}
\label{Fig06}
\end{figure}

In the PV diagram at offset 0$''$ along the major axis, we do not see a drastic velocity change with positions.
There seems to be a subtle difference in the line of sight velocity between the NE and SW ridges of the ring, and the entire PV diagram is marginally tilted.
To examine the velocity structure of the ring in more detail, we plotted in Figure~\ref{Fig06} the variations of peak brightness, peak velocity, and FWHM velocity width of the H29$\alpha$ line along the major (left) and minor (right) axes.
The values were measured by applying Gaussian fits to the line profiles on the PV diagrams.

%:%%%%% Fig06 schmatic image %%%%%
%%%%%  Fig06 %%%%%

The radial velocities along the minor axis are 26.3~km\,s$^{-1}$ and 6.0~km\,s$^{-1}$ at its SE ($-$0\farcs035) and NW ($+$0\farcs047) ridges, respectively.
The apparent ``smooth'' velocity gradient from the SE to NW ridge is naturally attributed to a stepwise velocity difference smeared by the beam, which has an HPBW similar to the ring width.
Because the velocity difference between the SE and NW ridges along the minor axis is caused either by expansion or contraction of the ring under our geometrical and kinematical assumptions, the velocity difference gives an expansion or contraction velocity of 13.2$\pm$1.0~km\,s$^{-1}$ at the radius of 706~au after correcting for the inclination angle, which is taken to be 50\degree.

The radial velocity along the major axis, which solely reflects the ring's rotation velocity under our assumption, varies from 17.2~km\,s$^{-1}$ to 21.4~km\,s$^{-1}$ as the position moves from the NE peak at $-$0\farcs073 to the SW peak at $+$0\farcs042.
The velocity difference corresponds to a rotation velocity of 2.7$\pm$1.0~km\,s$^{-1}$ at the radius of 642~au after correcting for the inclination angle.

\subsection{Expanding ring of ionized gas and its implication}\label{Expansion}

We have found a remarkable fact that the ionized ring of W49N:A2 has a predominant radial motion of 13.2$\pm$1.0~km\,s$^{-1}$, with a hint of rotation at 2.7$\pm$1.0~km\,s$^{-1}$, at $\sim$700~au in radius.
The ionizing source of A2 emits Lyman continuum photons equivalent to an O9 (ZAMS) star \citep{DePree2020}, which should have a mass of 20$\pm$7~M$_\odot$\footnote{Mean and standard deviation of mass for field O9 stars with the luminosity classes I, III and V.} \citep{Hohle2010}.
The ionized gas mass of A2 is 0.12~M$_\odot$ \citep{DePree2020}, which, together with the absence of molecular and dust emissions toward A2, implies that the total amount of mass within the ring cannot be much larger than 20~M$_\odot$.

It is impossible that this amount of mass gravitationally accelerates any gas to $\sim$13~km\,s$^{-1}$ at a distance of $R$ = 700~au from the center of A2.
We would need a total mass as large as 70~M$_\odot$ if a lump of gas freely falling toward the central star obtains the three-dimensional velocity of $\sim$13~km\,s$^{-1}$ at the radius.
Alternatively, we can derive a Keplerian mass of 5.3--5.7~M$_\odot$ from the inclination-corrected rotation velocity of 2.7~km\,s$^{-1}$ at a radius of 642--700~au.
This mass is significantly smaller than the expected mass of an O9 star.
We, therefore, conclude that the ring's observed radial motion cannot be gravitational infall or contraction.

Expansion, on the other hand, is a natural cause of motion.
At the beginning of HII region formation, ionized gas around a massive star is pushed outward due to the thermal pressure or magneto-hydrodynamical acceleration \citep{Commercon2022}. 
A radially expanding motion perpendicular to the outflow axis has, for example, been observed toward Orion Source~I \citep{Hirota2017}, where the expansion velocity of $\sim$10~km\,s$^{-1}$ was observed above and below the disk plane that is supposed to be driving a magneto-centrifugal outflow.

The current case of A2 has a linear scale 10 times larger than the case of Source~I, so we are unable to resolve the motion in the disk from that in the outflow.
However, the entire ionized ring, which could be a mixture of a disk and an outflow, shows expansion with a hint of rotation, while in the case of Source~I no radial motion was detected in the disk plane and rotation is dominant there.
In the case of W49N:A2, the dominant motion in its equatorial plane is expansion.

The small rotation velocity of the A2 ring is consistent with its expanding motion.
The observed rotation velocity at $R \sim$ 650~au is 2.7$\pm$1.0~km\,s$^{-1}$, which is significantly (2.5\,$\sigma$) smaller than the Kepler velocity of 5.2\,km\,s$^{-1}$ at the radius around a 20~M$_\odot$ star.
It may be the case that the gas at $R \sim$ 650~au has been transported from an inner radius, specifically from $R \sim$ 170~au.
The gas was originally in Keplerian rotation at the radius but was pushed out to the current radius by conserving its specific angular momentum of $\sim$1800~au\,km\,s$^{-1}$.
Such a phenomenon may occur when the accretion through the equatorial disk becomes hampered and reversed by the intense thermal or magneto-hydrodynamical pressure caused by the central star as its luminosity increases dramatically at the end of Kelvin-Helmholtz contraction \citep{Tanaka2016, Tanaka2017}.
 Although we cannot exclude the possibility of a small amount of accretion through the mid-plane of the disk even at this later stage of massive star formation, we did not find any evidence of accretion around the ionizing star of A2.

The ionized gas toward A2 shows broad H29$\alpha$ line widths of $\Delta V$ = 40--60~km\,s$^{-1}$, which may be attributed to a gas outflow with a large opening angle or to pressure broadening due to the high-density \citep{Brocklehurst1972, Kurtz2005}.
There is a marginal increase in line width inside the ring, seen on the upper panels of Figure~\ref{Fig06}.
The outflow there may have a higher velocity because it is accelerated in a region closer to the central star, or the outflow there is preferentially directed toward the polar direction.

\begin{figure}[tbh]
\includegraphics[bb=0 0 160 260, scale=0.45]{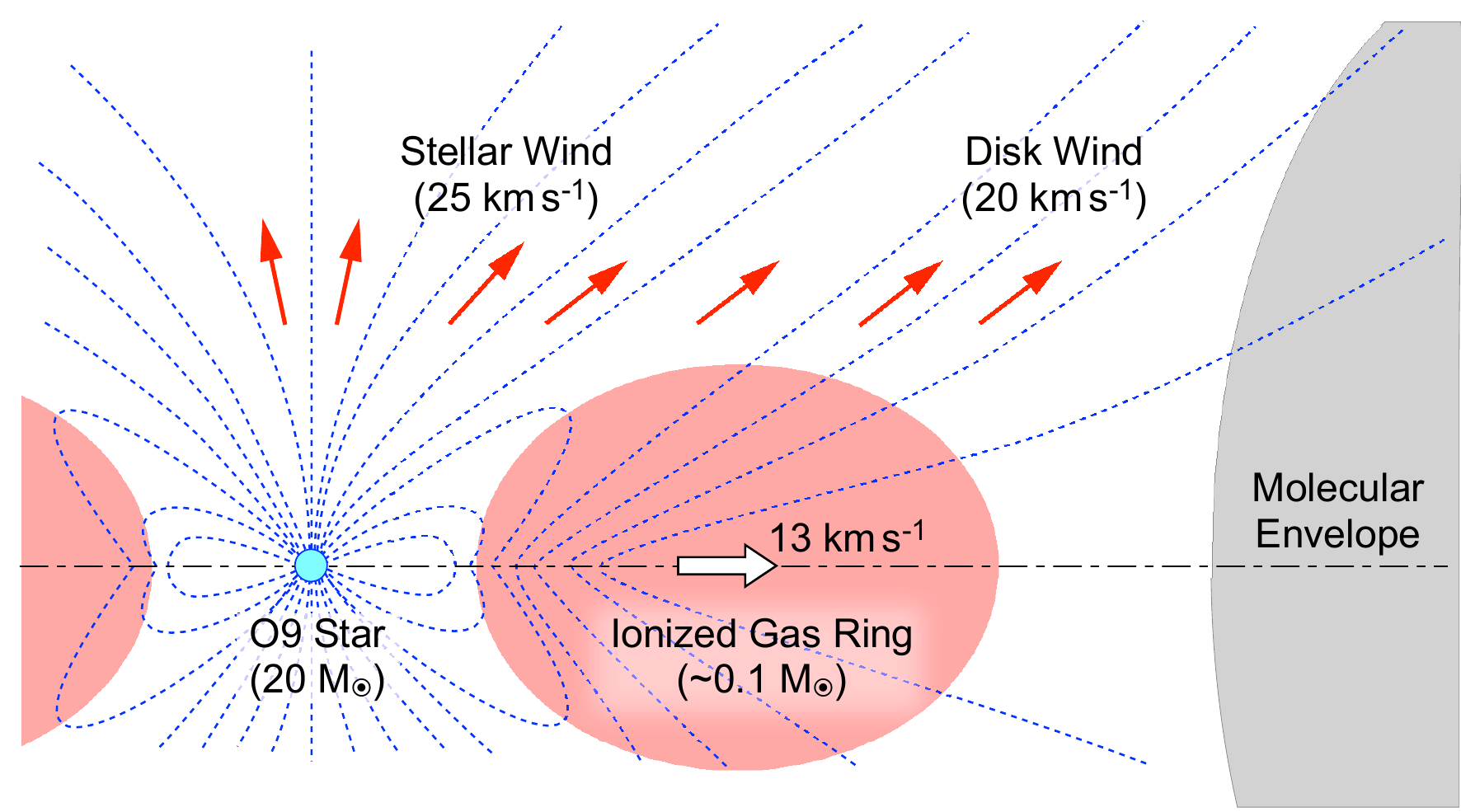}
\caption{A schematic drawing of W49N:A2. 
}
\label{Fig07}
\end{figure}

From these pieces of observational evidence, we conclude that the massive star ionizing the HCHII region W49N:A2 has finished its main accretion phase and is in the phase of disrupting its natal accretion disk.
W49N:A2 thus provides an example of how a massive star terminates its accretion at the end of its formation.
As time goes on, the ring will be fully evaporated by the stellar radiation.
The ionized outflow will expand by gradually decreasing its expansion velocity, and the HCHII region will eventually become a UCHII region.
Figure~\ref{Fig07} shows a schematic drawing of the observed features, where outflow velocities are derived from the H29$\alpha$ line widths deconvolved by its thermal widths at the assumed temperature of 10000~K \citep{DePree1997}.

%:%%%%% Fig07 schmatic image %%%%%
%%%%%  Fig07 %%%%

%: CONCLUSIONS
\section{Conclusions}\label{Conclusions}
We have presented ALMA archival data for 241--260~GHz continuum and  H29$\alpha$ recombination line observations toward the HCHII region W49N:A2 at a resolution of $\sim$0\farcs05. (600~au), confirming the presence of an ionized ring of $\sim$700~au in radius inclined by $\sim$50\degree.
The tilted ring has a prominent velocity difference between its NW and SE ridges along the minor axis, suggesting that it is expanding in the equatorial plane at a velocity of 13.2~km\,s$^{-1}$.

The ring also shows a hint of rotation at 2.7~km\,s$^{-1}$, which is significantly (2.5\,$\sigma$) smaller than the Kepler velocity of 5.2~km\,s$^{-1}$ at the ring radius around a 20~M$_\odot$ star.
This can be interpreted that the ring gas has been transported from the radius of $\sim$170~au by conserving its original specific angular momentum that it had there.
The ionized ring may thus be a remnant of the accretion disk that fed the O9 star, whose radiation or magnetic activities recently became so strong that the accretion was reversed.
 W49N:A2 provides a rare example of how a massive star terminates its accretion at the end of its formation and how an HCHII region transitions to a UCHII region. 

%: ACKNOWLEDGMENT
\begin{ack} 
ALMA is a partnership of ESO (representing its member states), NSF (USA), and NINS (Japan), together with NRC (Canada), NSC and ASIAA (Taiwan), and KASI (Republic of Korea), in cooperation with the Republic of Chile. 
The Joint ALMA Observatory is operated by ESO, AUI/NRAO, and NAOJ.
We used the ALMA archival data \#2018.1.00520.S (PI: D. Wilner), \#2018.1.00589.S (PI: R. Galv\'{a}n-Madrid), \#2016.1.00620.S (PI: A. Ginsburg), \#2015.1.01535.S (PI: B.~R. Alejandro), \#2017.1.00318.S (PI: R. Galv\'{a}n-Madrid), \#2017.1.01499.S (PI: F. Xiaoting).
\end{ack}

%: APPENDIX
%\begin{appendix}\label{APPENDIX}

%\end{appendix}

%: REFERENCES

\end{document}